\documentclass{emulateapj}
\newcommand{\zem}{z_{\mathrm{em}}}
\newcommand{\zabs}{z_{\mathrm{abs}}}
\newcommand{\delv}{v_{\mathrm{ej}}}
\newcommand{\aox}{\alpha_{\mathrm{ox}}}
\newcommand{\aopt}{\alpha_{\mathrm{o}}}
\newcommand{\axray}{\alpha_{\mathrm{x}}}
\newcommand{\fnu}{F_{\mathrm{\nu}}}
\newcommand{\kms}{km~s$^{-1}$}
\newcommand{\flamunit}{\mathrm{erg~cm}^{-2}~\mathrm{s}^{-1}~\mathrm{\AA}^{-1}}
\newcommand{\fnuunit}{\mathrm{erg~cm}^{-2}~\mathrm{s}^{-1}~\mathrm{Hz}^{-1}}
\newcommand{\lumnuunit}{\mathrm{erg~s}^{-1}~\mathrm{Hz}^{-1}}
\newcommand{\lumunit}{\mathrm{erg~s}^{-1}}
\newcommand{\HI}{H\,{\sc i}}
\newcommand{\CI}{C\,{\sc i}}
\newcommand{\CII}{C\,{\sc ii}}
\newcommand{\CIV}{C\,{\sc iv}}
\newcommand{\NI}{N\,{\sc i}}
\newcommand{\NV}{N\,{\sc v}}
\newcommand{\OI}{O\,{\sc i}}
\newcommand{\OVI}{O\,{\sc vi}}
\newcommand{\MgI}{Mg\,{\sc i}}
\newcommand{\AlII}{Al\,{\sc ii}}

\newcommand{\SiII}{Si\,{\sc ii}}
\newcommand{\SiIII}{Si\,{\sc iii}}
\newcommand{\SiIV}{Si\,{\sc iv}}
\newcommand{\PII}{P\,{\sc ii}}
\newcommand{\PIII}{P\,{\sc iii}}
\newcommand{\SI}{S\,{\sc i}}
\newcommand{\SII}{S\,{\sc ii}}
\newcommand{\SIII}{S\,{\sc iii}}
\newcommand{\MnII}{Mn\,{\sc ii}}
\newcommand{\FeII}{Fe\,{\sc ii}}

\newcommand{\NiII}{Ni\,{\sc ii}}
\newcommand{\CuII}{Cu\,{\sc ii}}
\newcommand{\Lya}{Ly\,$\alpha$}
\newcommand{\Lyb}{Ly\,$\beta$}
\newcommand{\Hb}{{H}\,\beta}
\newcommand{\CIVdblt}{C\,{\sc iv}~$\lambda\lambda 1548.2, 1550.7$}
\newcommand{\NVdblt}{N\,{\sc v}~$\lambda\lambda 1238.8, 1242.8$}
\newcommand{\OVIdblt}{O\,{\sc vi}~$\lambda\lambda 1031.9, 1037.6$}
\newcommand{\SiIVdblt}{Si\,{\sc iv}~$\lambda\lambda1393.8, 1402.8$}
\begin{document}

%\received{31 May 2002}
\accepted{14 Aug 2003}
%\journalid{number}{date month year}
%\articleid{number}{number}

%\slugcomment{in preparation for {\it The Astrophysical Journal};Version 4.4}
\shortauthors{Ganguly \etal}
\shorttitle{RX~J$1230.8+0115$}

%\title{\Large\bf An Intrinsic Absorption Complex Toward {RX~J$\mathbf{1230.8+0115}$}: Geometry and Photoionization Conditions}

\title{An Intrinsic Absorption Complex Toward {RX~J$1230.8+0115$}: Geometry and Photoionization Conditions}

\author{Rajib Ganguly \altaffilmark{1,2}, Joseph Masiero \altaffilmark{1}, Jane C. Charlton
\altaffilmark{1,3}, Kenneth R. Sembach \altaffilmark{2}}

\altaffiltext{1}{Department of Astronomy and Astrophysics, 525
Davey Lab, University Park, The Pennsylvania State University,
State College, PA 16802}

\altaffiltext{2}{The Space Telescope Science Institute, 3700 San
Martin Drive, Baltimore, MD 21218}

\altaffiltext{3}{Center for Gravitational Physics and Geometry,
The Pennsylvania State University}

\begin{abstract}
We present HST/STIS and FUSE spectra of the quasar
{RX~J$1230.8+0115$} ($V=14.4, z=0.117$). In addition to Galactic,
Virgo, and intervening absorption, this quasar is host to a
remarkable intrinsic absorption complex. Four narrow absorption
line systems, strong in {\CIV}, {\NV}, and {\OVI}, lie within
{$5000$~\kms} of the QSO redshift. Three of the systems appear to
be line-locked, two in {\NV}, and two in {\OVI}, with the common
system residing in between the other two (in velocity). All three
systems show signs of an intrinsic origin -- smooth wind-like
profiles, high ionization, and partial coverage of the central
engine. The fourth system, which appears at the systemic redshift
of the QSO, may originate from host galaxy or intervening gas.
Photoionization analyses imply column densities in the range
{$19.1 < \log N$(H) $ < 21$} and ionization parameters in the
range {$-1.3 < \log U <  0.3$}. Revisiting the issue of
line-locking, we discuss a possible model in the context of the
accretion-disk/wind scenario and point out several issues that
remain for future simulations and observations.
\end{abstract}


\section{Introduction}
\pagestyle{myheadings}

Narrow absorption lines due to gas that is truly intrinsic to the
central engines of quasi-stellar objects (QSOs) have recently been
recognized as ubiquitous {\citep{rich99}}. They are a powerful
diagnostic of the physical conditions of the gas {\citep[e.g.,
][]{ham97a,cmm98,ham00,kra01}} since they arise from a wide range
of ionization conditions. Still, the origins of this gas and its
relationship to other regions (e.g., the broad emission line
region; hereafter, the BLR) is uncertain. There are indications
that many of the {\it associated} absorption lines (narrow
absorption lines that arise within {$5,000$~\kms} of the QSO
emission redshift) are indeed connected to the BLR
{\citep{gan01a}}, and to the X-ray ``warm'' absorbers {\cite[e.g.,
][]{mew95,brandt}}. Strong [$W_{\mathrm{r}}\gtrsim1.5$~\AA]
associated absorbers seem to appear preferentially in
optically-faint, steep-spectrum, radio-loud objects
{\citep{mj87,foltz88,ald94}}, while ``high ejection velocity''
absorbers tend to appear in the spectra of radio-quiet and
flat-spectrum objects {\citep{rich99}}.

Another puzzle regarding intrinsic narrow absorption is the origin
of complexes of intrinsic lines. Here, a complex of absorption
lines means a group of narrow absorption line systems, which are
well separated in velocity, along a particular line of sight where
the local redshift path density is much larger than expected from
randomly distributed intergalactic gas. This is observationally
distinct from the substructure that nearly all absorption line
systems (intrinsic or intervening) exhibit at high spectral
resolution. There are many examples of intrinsic absorption
systems (with single profiles) which show substructure
{\cite[e.g., ][]{kor93,cren99,gan99}}. While several examples of
absorption lines complexes now exist in the literature
{\citep[e.g., ][]{foltz87,dekool01,gan01b,rich02,misawa03}}, their
origin has not yet been fully explored. Some fraction of complexes
may represent cosmological ``superstructure,'' but others result
from gas intrinsic to the background quasar. While a small
fraction of sightlines are expected to contain excesses of
absorption line systems, the incidence of absorption line
complexes has not yet been sufficiently constrained to infer the
relative contributions of these two scenarios. In the cases where
such complexes are due to intrinsic gas, they represent an
important constraint on unification models designed to explain
intrinsic absorption. Under the intrinsic hypothesis, it is
unclear how line complexes relate to broad absorption lines, to
single intrinsic narrow absorption lines with extensive
substructure, or even to the properties of the ``host'' QSO. In
this paper, we report an intrinsic absorption line complex toward
the QSO {RX~J$1230.8+0115$} ($V=14.42$, $\zem=0.117$, $\nu
L_{\nu}$(2500\,\AA)$=2.8\times10^{44}~\lumunit$).

{RX~J$1230.8+0115$} was first detected by the ROSAT All-Sky Survey
{\citep*{rmh98}}. It is located {$54'$} from 3C~273 on the sky and
is one of the brightest QSOs in the optical band. It hosts a
complex of four absorption lines (detected in {\HI}, {\CIV},
{\NV}, and {\OVI}) in the ejection velocity range {$-4600 < \delv
< -215$~km s$^{-1}$}, three of which are likely to have an
intrinsic origin. In \S\ref{sec:rxjdata}, we present high
resolution {\it Hubble Space Telescope} (HST) and intermediate
resolution {\it Far Ultraviolet Spectroscopic Explorer} (FUSE)
spectra of {RX~J$1230.8+0115$}, and a description of the complex.
In \S\ref{sec:rxjproof}, we show that the complex has an intrinsic
origin. In \S\ref{sec:rxjphoto}, we characterize the photoionizing
spectrum and present results from Cloudy {\citep{hazy}}
simulations to constrain the physical conditions of the gas. We
summarize our results in \S\ref{sec:rxjresults}. Finally, in
\S\ref{sec:rxjdiscuss}, we discuss the implications of our results
in the broader picture of intrinsic absorbers and intrinsic
absorption complexes and point out avenues (both theoretical and
observational) for further research.

\section{Data}
\label{sec:rxjdata}

{RX~J$1230.8+0115$} has been observed several times by HST for the
purpose of examining {\Lya} absorption in the local universe
{\citep*{imp99,pss00a,pss00b}}. More recently, {RX~J$1230.8+0115$}
was observed for 27.2~ksec in January 1999 with the {\it Space
Telescope Imaging Spectrograph} (STIS) using the E140M echelle,
{$0.\!''2 \times 0.\!''06$} slit, and the FUV-MAMA detector (under
proposal 7737 by Michael Rauch). The resolving power of this
configuration is {$R\sim45,600$} ({6.6~\kms} velocity resolution)
with two pixels per resolution element. The reduction of the raw
data and calibration of the reduced data followed the standard
HST/STIS pipeline software (the {\tt CALSTIS} package in
IRAF\footnote{IRAF is distributed by the National Optical
Astronomy Observatories, which are operated by AURA, Inc., under
contract to the NSF.}). According to the {\tt CALSTIS}
documentation {\citep{calstis}}, the absolute wavelength
calibration is good to within a pixel ({0.015~\AA} at {1425~\AA},
the central wavelength), and the absolute photometry is good to
8\% ($\sim 2 \times 10^{-15}~\flamunit$). The fully reduced and
calibrated spectrum, covering the wavelength range
{1178.2--1723.8~\AA}, is presented in Fig.~\ref{fig:rxjstisplot}.
The data have been resampled to combine overlapping regions of the
echelle orders. In the figure, the spectra have also been rebinned
($2\rightarrow1$) to resolution element, not pixel, samples.
Superimposed on the data is the effective continuum (continuum
{$+$} emission lines) fit. The lower trace in each panel is the
error spectrum.

{RX~J$1230+0115$} was observed by FUSE on 20 Jun 2000 for a total
exposure time of 4~ksec.  The four FUSE channels were co-aligned
during the observation, and signal was detected across the full
FUSE bandpass (905-1187 A). We retrieved the raw data from the
FUSE archive (observation set P1019001) and calibrated the data
for each detector segment using the latest version of {\tt
CALFUSE} (v2.1.6).  {\tt CALFUSE} screens the photon event lists
for valid data, performs geometrical distortion corrections, and
calibrates the one dimensional spectra extracted from the two
dimensional spectral images.  After spectral extraction, we
checked the heliocentric wavelength calibration of each segment by
comparing common spectral regions recorded in different data
segments.

For this analysis, we restrict our attention to data from the
LiF1b and LiF2a channels which cover the wavelength range
1086.4--1184.8 (shown in Fig.~\ref{fig:rxjfuseplot}) since the
data from the SiC1 and SiC2 channels have low {$S/N$}. We binned
the data into 5-pixel bins to improve {$S/N$}, while maintaining
the full velocity resolution of the data. (The raw data are
oversampled at 10--12 pixels per resolution element.) We estimate
{$S/N = 4 (5)$} per {20~\kms} resolution element at {1150~\AA} for
the LiF1b (LiF2a) channel.

We note here that, while {RX~J$1230.8+0115$} is an X-ray selected
quasar, there are no X-ray observations that are of sufficient
quality to extract a spectrum. A search in the {\sc heasarc}
archive reveals only one ``direct'' observation - that of the
ROSAT All-Sky Survey. While there are several pointed observations
of 3C~273, {RX~J$1230.8+0115$} is at the edge of these fields
where the point-spread function is large, and the there are
insufficient counts to create a spectrum. An observation with
Chandra or XMM-Newton would be very revealing in regards to a
possible ``warm absorber'' phase.

\subsection{Line Identifications and Measurements}

Our method for detecting absorption features followed the
prescription from {\citet*{ltw87}}. This method creates a smoothed
equivalent width spectrum from both the normalized flux and the
normalized errors (the sigma). A pixel records absorbed flux if
the smoothed equivalent width ($W_{\mathrm{i}}$) exceeds the error
($\sigma_{\mathrm{i}}$) by a user-defined threshold (i.e.,
{$W_{\mathrm{i}}/\sigma_{\mathrm{i}} > n$}). In this case, we set
our detection limit to {$1.5\sigma$}.

At this low confidence level, our completeness limit for detecting
real features of a given equivalent width is improved, but the
number of false detections is increased. We are {95\%} complete
down to a limiting equivalent width of {$0.10$~\AA} at
{$1.5\sigma$} confidence. This limit increases to {$0.31$~\AA} at
{$4.5\sigma$} confidence. To limit the false detection of an
absorption system, we require the detection of multiple species --
a spectral doublet ({\NVdblt} or {\CIVdblt}) and {\Lya}. To
characterize the false detection rate of this approach, we
simulated 1000 noise realizations using the error spectrum from
the STIS-E140M observation and looked for sets of features that
(falsely) satisfied the {\NVdblt$+$\Lya} detection criterion. In
these realizations, 3517 features were detected at {$1.5\sigma$}
confidence that could potentially yield an absorption system in
the redshift range {$0 < z < 0.1358$} in {\NVdblt} and {\Lya}.
(The upper redshift limit arises from the canonical upper limit
for associated systems: {5000~\kms} beyond the quasar emission
redshift.) Since our criterion for the selection of a system
requires the detection of a set of three lines, 3517 features
translates to {$7\times10^9$} potential systems - the number of
ways of choosing a set of 3 lines from a sample of 3517. Of the
{$7\times10^9$} sets of features, {$5\times10^{5}$} satisfied the
{\NV~doublet$+$\Lya} spacings and therefore would have been
identified (falsely) as an absorption system. This implies a false
detection rate of {$5\times10^5/7\times10^9 = 7\times10^{-5}$}.
Thus, in spite of the low ($1.5\sigma$) confidence for
``detecting'' features, our confidence threshold for {\it
identifying a system} is 99.993\% (effectively $4\sigma$).

To identify other features with absorption systems, we compiled a
list of resonant transitions covered by the spectrum for any
possible absorber between the Galaxy and the QSO. With both the
STIS-E140M spectrum and FUSE LiF spectra, the effective observed
wavelength range is 1128.3--1723.8~\AA. (Due to the redshift of
the quasar, we include in our line-list transitions down to a
rest-frame wavelength of 1010.1~\AA.) After identifying Galactic
lines, we culled the list of detected features using the
aforementioned criteria with the {\NVdblt} and {\CIVdblt}
doublets. As a further constraint, the resulting candidate systems
were checked by hand to ensure that doublets had similar kinematic
structures. In the end, we identified six metal-line systems, four
of which lie within {$5000$~\kms} of the QSO emission redshift.
These four systems are described in the next section. [One of the
other two systems is absorption by the Virgo Southern Extension.
This system has been examined by {\citet*{gan03a} and
{\citet{ros03}}.] In Fig.~\ref{fig:rxjvelocity}, we show the
detected transitions of the complex from {\HI}, {\CIV}, {\NV}, and
{\OVI} aligned in velocity with the zero-point at {$\zem=0.117$}.

\subsection{Associated Absorption Systems} \label{sec:rxjmetals}

\begin{itemize}
\item[$z =$] 0.1000: (System A in Fig.~\ref{fig:rxjvelocity}) This
is a weak system that is detected in {\HI}, {\CIV}, {\NV}, and
{\OVI}. The {\OVIdblt} transitions in this system are apparently
line-locked with the {\Lyb} and {\OVI~$\lambda1031.9$} transitions
from System B at {$\zabs=0.1058$}. The {\OVI~$\lambda1031.9$} line
is blended with Galactic absorption by the
{\NI$\lambda\lambda\lambda1134.2,1134.4,1135$} triplet.

\item[$z =$] 0.1058: (System B in Fig.~\ref{fig:rxjvelocity}) This
system is detected in {\HI}, {\CIV}, {\NV}, and {\OVI}. This
system is apparently line-locked in the {\NVdblt} doublet with
that from the {$z = 0.1093$} system (System C). The system is also
apparently locked with the {$z=0.1000$} system (System A) in the
{\OVIdblt} doublet (as mentioned above). The {\CIV} doublet for
this system is self-blended (with a width larger than the doublet
separation) so this is not a classic narrow absorption line
system. The smooth troughs of all the transitions are
characteristic of wind kinematics. In addition, the
{\NV$\lambda1232.8$}, and {\HI~\Lyb} lines are blended with the
{\NiII$\lambda1370.1$} and
{\NI$\lambda\lambda\lambda1134.2,1134.4,1135$} Galactic
absorption, respectively. This type of system has been referred to
as a mini-BAL {\citep*[e.g.,][]{turn88b,bhs97,cssg99}}, a
relatively narrow version of a BAL.

\item[$z =$] 0.1093: (System C in Fig.~\ref{fig:rxjvelocity}) This
system is detected in {\HI}, {\CIV}, {\NV}, and {\OVI}. The
{\NV$\lambda1238.8$} transition for this system is blended with
the {\NV$\lambda1242.8$} transition of System B (as mentioned
above) and the {\OVI~$\lambda1031.9$} line is blended with
Galactic {\FeII~$\lambda1144.9$} absorption. The profiles of this
system are also smooth, indicative of an outflowing-wind origin.
For clarification, we note that the line at 1374.2\,\AA,
misidentified by {\citet{imp99}} as a {\Lya} absorber at {$\zabs =
0.1301$} from a low resolution HST/GHRS-G140L spectrum, is
actually the blend between \NV$\lambda1242.8$ at {$\zabs =
0.1058$} and \NV$\lambda1238.8$ at {$\zabs = 0.1093$}. The
blending of the {\NV} doublet is clear from the higher resolution
spectrum presented here.

\item[$z =$] 0.1162: (System D in Fig.~\ref{fig:rxjvelocity}) This
system is detected in {\HI}, {\NV}, and {\OVI}. The
{\OVI~$\lambda1031.9$} line is blended with Galactic
{\PII~$\lambda1152.8$}. The redshift actually places this system
redward of the {\NV} emission line peak, although the velocity
matches the QSO systemic redshift (as measured using Balmer
lines). The kinematic structure of this system clearly motivates
the idea that the absorption arises from a clumpy medium (at least
along the line of sight). Another clarifying note: The
{\NV$\lambda1242$} feature for this system was misidentified by
{\citet{imp99}} as a {\Lya} absorber at {$\zabs = 0.1419$}.

\end{itemize}

\section{Are They Truly Intrinsic?}
\label{sec:rxjproof}

\subsection{Circumstantial Evidence}

We examine three avenues of circumstantial evidence that point
toward an intrinsic origin of the absorbing gas in this complex.
In order of increasing reliability, we discuss the probability of
detecting these systems in the redshift path, the probability for
chance superposition of lines (``line locking''), and the high
ionization state of the absorbers.

\subsubsection{Redshift Path Density of Absorbers}

The first clue that these absorbers have an intrinsic origin comes
from the low probability that they arise from uncorrelated gas
(i.e., intergalactic clouds), as expected if the absorbers were
cosmologically distributed. In this sightline, six metal-line
systems are detected in the {\CIVdblt} doublet and four in the
{\NVdblt} doublet down to a limiting equivalent of {$W>10$~m\AA}
over a redshift path of {$\delta z = 0.1162$}. According to the
HST Quasar Absorption Line Key Project, on average we expect to
detect 0.11 {\CIV} absorbers and 0.01 {\NV} absorbers in this
redshift path down to a 95\% completeness limit of
{1~\AA}.\footnote{We have computed this limit using the Key
Project spectra, which were generously donated fully and uniformly
reduced and calibrated with continuum fits by B. Jannuzi, S.
Kirhakos, and D. Schneider. The Key Project observations,
reductions, and line lists are detailed in {\citet{kpi}},
{\citet{kpii}}, {\citet{kpvii}}, and {\citet{kpxiii}}.} At this
same equivalent width limit, we detect only 1 {\CIV} system and 2
{\NV} systems. The Poisson probability of finding 1 {\CIV} system
(2 {\NV} systems) when only 0.11 (0.01) are expected is 0.10
($5\times10^{-5}$). Based on the statistics of {\NV} absorbers, it
is probable that the absorbers are correlated. In the next
section, we examine the likelihood of observing three apparently
line-locked metal-line systems if the systems are uncorrelated.

\subsubsection{Apparent Line-Locking of Systems}

We performed a Monte Carlo simulation to determine the chance
probability of detecting apparent line-locking among four {\NV}
systems placed randomly over a redshift path. Two systems are
taken to be apparently line-locked if the velocity separation
matches that of the {\NVdblt} doublet, {964.5~\kms}, to 90\%
confidence. (In the simulations, we used the velocity widths of
the four associated systems: 40, 137, 65, 63~\kms.) For a redshift
path of {$\Delta z = 0.0333$} (corresponding to the velocity
window for ``associated'' systems), the probability of a chance
superposition between two systems is 0.070. (The probability is
similar for the {\OVI} doublet separation.) The binomial
probability of finding two apparently line-locked pairs, from a
sample of 6 pairs (4 absorption systems yields 6 pairs) is 0.055.
Thus, it is unlikely that the apparent line-locking of systems in
this sightline is due to random chance.

\subsubsection{High Ionization State}

High ionization structure is used (usually in tandem with other
evidence) as a sign of an intrinsic origin for absorbers, since
the gas is subject to photoionization by the quasar continuum
{\citep[e.g., ][]{ham95,bs97}}. It is readily seen (before
detailed photoionization modelling) that the absorbers are highly
ionized. There is no Lyman limit detected in the FUSE spectrum, so
all the {\HI} column densities must be
{$<10^{17.2}~\mathrm{cm}^{-2}$} assuming no dilution by unocculted
flux. The {\NV} and {\CIV} absorption profiles are very strong
compared to their {\Lya} profiles; in fact, {\NV} appears to be
saturated in the case of system B. Finally, while the
{\SiIII$\lambda1206.5$} and {\SiIVdblt} lines are covered by the
STIS-E140M spectrum, they are not detected down to limiting
(3$\sigma$) equivalent widths of {$\sim0.01$~\AA} (for \SiIII) and
{$\sim0.03$~\AA} (for \SiIV). These limits correspond to column
density limits of {$N$(\SiIII)$\lesssim
10^{11.7}~\mathrm{cm}^{-2}$} and {$N$(\SiIV)$\lesssim
10^{12.5}~\mathrm{cm}^{-2}$} assuming there is no dilution by
unocculted flux.

\subsection{Direct Evidence}
Since the spectral resolution of the STIS-E140M observation is
good enough (FWHM$\approx$6--7~\kms) to fully resolve the
kinematics of the profiles, it is possible to apply a test to
determine if the associated absorbers are truly intrinsic. Using
the well-separated {\NVdblt} and {\CIVdblt} doublets, one can
infer the fraction of the background flux incident on the
absorber. Normally this is done by using the flux in corresponding
pixels of the two doublet profiles. This can be done as a function
of velocity {\citep[e.g.,][]{akd02}} by computing the coverage
fraction across the entire profile, or as a function of kinematic
component {\citep[e.g.,][]{ham97a}} by using the flux in the cores
of the components. The premise behind this approach is that
profiles are diluted by flux that is not incident on the absorber.
The apparent attenuation of the lines is given by:
\begin{equation}
I(v) = 1 - C(v) + C(v) e^{-\tau(v)}. \label{eq:atten}
\end{equation}
Invoking the atomic physics constraint that the true optical
depths of the spectral doublet transitions have a 2:1 ratio yields
the standard solution for the coverage fraction as a function of
velocity:
\begin{equation}
C(v) = {{(I_{\mathrm{w}}(v) - 1)^2} \over {I_{\mathrm{s}}(v) - 2
I_{\mathrm{w}}(v) + 1}}, \label{eq:parcov}
\end{equation}
where {$I_{\mathrm{s}}$} and {$I_{\mathrm{w}}$} are the normalized
flux profiles of the stronger and weaker transitions,
respectively. Due to instrumental effects, a pixel-by-pixel
computation (i.e, computations across the entire profile) is not
trustworthy in the wings of profiles even if the profiles are well
resolved {\citep{gan99}}. In passing, we also note that the
correct geometric interpretation of the coverage fraction is not
clear. (For example, light scattered around the absorbing gas can
produce coverage fractions less than unity.)

In the case of this complex, it is only possible to apply a
pixel-by-pixel approach to three doublets -- the {\NVdblt} doublet
for systems A and D, and the {\CIVdblt} doublet for system C. No
doublets for system B are sufficiently ``clean'' for this
approach, and the FUSE spectra are not off sufficient quality to
apply this to the {\OVIdblt} profiles. In
Fig.~\ref{fig:rxjparcov}, we show two panels for each system that
test for the existence of partial coverage in these doublet
profiles. The top panels show the apparent column densities
{\citep[][hereafter, ACD]{ss91}} of the doublet transitions (and
differences therein). When profiles are resolved and apparently
unsaturated, the ACD profiles from both transitions should
coincide. The signature of partial coverage is that the weaker
transition will imply a larger ACD than the stronger transition,
although this can also result from unresolved saturation. In the
bottom panels, we show the normalized flux profile from the
stronger transition and the pixel-by-pixel coverage fraction (from
eq.~\ref{eq:parcov}). Coverage fractions are plotted as per the
criteria of {\citet{gan99}}.

The {\NV} doublet for system A clearly shows evidence for partial
coverage from both the ACD and pixel-by-pixel coverage fraction
profiles. The {\CIV} doublet toward system C shows no evidence for
partial coverage. System D shows marginal evidence at the core of
the strongest component from the ACD plot, but not in the coverage
fraction plot. This may be unresolved saturation from a narrow
component.

In the interest of extracting column densities for all absorption
components, we employed a profile fitting technique that separately
accounts for the partial coverage fractions and column densities.  We
model the profiles as arising from discrete, Gaussian-broadened
components which only partly occult the backround source. In the limit
where this is applicable, each component givesrise to an effective
attenuation factor given by Eq.~\ref{eq:atten} and the observed
profile is given by the product of these factors.
\begin{equation}
I(\lambda) = I_{\mathrm{o}}(\lambda) \prod_{i=1}^{n}
\prod_{j=1}^{m(i)} \left [ 1 - C_{\mathrm{i}} + C_{\mathrm{i}}
e^{-\tau_{\lambda}(z_{\mathrm{i}}; N_{\mathrm{ij}},
b_{\mathrm{ij}})} \right ], \label{eq:cloudattenuation}
\end{equation}
where {$n$} is the number of absorption components, {$m(i)$} is
the number of ionic species associated with component {$i$},
{$C_{\mathrm{i}}$}, and {$z_{\mathrm{i}}$} are the partial
coverage fraction and redshift of the {$i^{\mathrm{th}}$}
component, and {$N_{\mathrm{i}}$}, and {$b_{\mathrm{ij}}$} are the
column density, and velocity width of the {$j^{\mathrm{th}}$}
species of component {$i$}. In the case where the entire flux from
the quasar is incident on all the absorbers (e.g., an intervening
absorber), this reduces to the exponential of the sum of optical
depths, {$I(\lambda) = I_{\mathrm{o}}(\lambda) \exp
[-\sum\tau(\lambda)]$}. In the case where the absorbers are
optically thick at the observed wavelength, this reduces to the
product of non-incident fractions,
{$I(\lambda)=I_{\mathrm{o}}(\lambda) = \prod
(1-C_{\mathrm{i}})$}.\footnote{In the appendix, we show that this
model provides results consistent with the direct inversion
techniques, and explore the geometric implications of the
prescription in a subsequent paper. We note here that this
consistency may imply an important issue in the interpretation of
the coverage fraction. The coverage fraction is the fraction of
{\it available}, not {\it total} sightlines occulted by the
absorber.}

In the application to this complex, we first fit the absorption
``by-hand'', using the minimum number of components possible and
the number of troughs in each system as an indication of the
number of components -- one component for system A, four for
system B, three for C, and five for D. The result was run through
the Numerical Recipes Levenberg-Marquardt {$\chi^2$}-minimizing
routines {\sc mrqmin} and {\sc mrqcof} {\citep{nrpress}} for
optimization and error analysis. The optimization procedure
minimizes the {$\chi^2$} statistic and only retains components
that significantly reduce {$\chi^2$} to 80\% confidence (this is
assessed using the F-test for {$\chi^2$} significance). We
constrain the coverage fraction and column density separately for
each detected ion. Coverage fractions are allowed to vary on the
(physically meaningful) range [0,1]. Column densities are allowed
to vary such that the optical depth in the strongest transition
lies in the range [0,3]. Beyond this range, the component is
deemed ``saturated'' and only a lower limit on the column density
is reported. We tie the redshift and velocity width of each the
absorption component across all detected ions and assume Gaussian
line-broadening. In other words, the redshift and velocity width
were allowed to vary, but not separately from ion to ion. (As a
first step, we separately constrained the contributions of thermal
and non-thermal broadening to the line widths; the thermal
contribution was found to be negligible.) We accounted for the
Galactic {\NiII$\lambda1370$} absorption, which blends with the
{\NV$\lambda1238$} absorption in System B, by simultaneously
fitting the {\NiII$\lambda1317$} line which has similar strength.
The optimization procedure threw out one component from system B
(since the absorption could be explained through the Galactic
{\NiII}) and one component from system C; the results appear in
Table~\ref{tab:rxjintrin}. We list the component name (column 1),
redshift (column 2), velocity width (column 3), {\HI} column
density (column 4), {\CIV} column density and coverage fraction
(columns 5 and 6, respectively), and {\NV} column density and
coverage fraction (columns 7 and 8, respectively). The {$1\sigma$}
fitting uncertainties for each parameter are listed beneath each
value.

We overlay the fit and the component contributions in
Fig.~\ref{fig:rxjvelocity}. For ions where only a single
transition is detected, the measurement of the coverage fraction
and column density are degenerate unless the transition is
saturated. In these cases, ``measured'' values are to be taken
lightly. Since the FUSE spectrum is neither of high spectral
resolution, nor of high signal-to-noise, we were unable to
reasonably (and directly) constrain the column densities and
coverage fractions of {\OVI} and {\HI} (using the higher Lyman
series lines). Since the {\Lya} transition for {\HI} is covered by
the STIS echelle spectrum, we obtained a lower limit on the {\HI}
column density by fixing the coverage fraction at unity.

Systems A, B, and C appear to require partial coverage to fit the
{\NV} and {\CIV} doublet profiles. Thus, it is likely that the
absorbing gas is close to the central engine. We note that the
partial coverage fractions reported for the ``clean'' systems
(described above) are consistent with those yielded by the direct
inversion. We also note that the absorption troughs are
sufficiently deep to imply that both continuum and broad emission
line photons are absorbed. Thus, the absorbers must either lie
beyond or be co-spatial with the broad line region. System D does
not require any partial coverage. Given the clumpiness of the
absorption and the closeness to the systemic redshift, it is
possible that this is absorption by the host galaxy. In the next
section, we use the measurements of the column densities and {\tt
Cloudy} simulations to constrain the physical conditions of the
absorbers.

\section{Physical Conditions of the Intrinsic Absorption Complex}
\label{sec:rxjphoto}

To constrain the physical conditions of the components, we applied
{\tt Cloudy} photoionization models to compare the {\CIV} and
{\NV} column densities.

\subsection{Characterization of the Ionizing Spectrum}

In our efforts to perform detailed photionization models of the
absorbers, we used the available X-ray, ultraviolet, and radio
data to characterize the photoionizing spectrum from the QSO. The
QSO is not detected by the FIRST survey {\citep{first}} down to a
limiting flux of 3.45~mJy, implying a 5\,GHz luminosity density
{$L_{\nu} < 10^{30.5}\,\lumnuunit$}. The QSO is radio-quiet and we
assume an X-ray energy index {$\axray=-1.69$} {\citep{laor97}}.
The ROSAT All-Sky Survey detection of the QSO reports a count rate
of {$0.04\pm0.01$\,counts~s$^{-1}$} in the 0.1--2\,keV band. Using
the {\tt HEASARC WebPIMMS} count rate-to-flux converter, we
calculate {$\fnu(2~\mathrm{keV}) = (3.0\pm1.0)\times
10^{-31}~\fnuunit$} assuming a power law model with Galactic
absorption [{$N(\mathrm{H})=1.83\times10^{20}~\mathrm{cm}^{-2}$}
{\citep{dl90}}]. From a power-law fit to the STIS-E140M spectrum,
excluding emission lines, we constrain the ultraviolet energy
index to be {$\aopt = -1.0$}, with {$\fnu(2500~\mathrm{\AA}) = 8.4
\times 10^{-27}~\fnuunit$}. From this, we infer {$\aox = -1.71$}.
For the ionizing spectrum, we used the {\tt Cloudy} AGN spectrum
with the following parameters: {$T=150,000~{\mathrm{K}}$}~(the
default), {$\aox = -1.71$}, {$\aopt = -1.0$}, {$\axray = -1.69$}.

We ran a two-dimensional grid of {\tt Cloudy} models stepping
through total hydrogen column densities [$N(\mathrm{H})$] and
ionization parameters ($U$). (The ionization parameter is defined
as the ratio of the number density of hydrogen ionizing photons to
the total hydrogen number density.) For each model, we chose a
plane-parallel slab geometry with solar abundances, and a nominal
hydrogen space density of {$3000~\mathrm{cm}^{-3}$} {\citep[e.g.,
][]{ham97a}}. The ionizing continuum was normalized by the
specification of the ionization parameter. For a given ionization
parameter, the space density of the gas and the source-absorber
distance are degenerate through the luminosity of the quasar ($U
\propto L/nr^2$). For an ionization parameter of {$U=0.1$}, the
effective distance is {$r\sim270$~pc}. Through the degeneracy, the
absorbers may be closer to the central engine if they are denser
than the assumed value. Since we do not have an independent
measure of the density (e.g., from detections of excited state
lines or from time variability of absorption profiles), we cannot
use the photoionization models to constrain either the
source-absorber distance, or the absorber thickness. Through
simulations, we have verified that the inferred ionization
parameters and total hydrogen column density are largely
independent on the choice of space density (over at least three
decades). For a given {$U$} and {$N$(H)}, increasing the space
density of an absorber will bring it closer to the ionizing source
and also make it smaller (i.e., the thickness decreases to match
the column density).

\subsection{Total Column Densities and Ionization Parameters}

Using the measured {\CIV} and {\NV} column densities (which are
automatically corrected for the covering fraction by the fitter)
and the {\HI} column density limit, we constructed column density
isopleths for each ion in {$N(\mathrm{H})-U$} space. In principle,
the intersection of these isopleths uniquely specifies both
{$N(\mathrm{H})$} and {$U$} for each absorption component. In
Fig.~\ref{fig:rxjionize}, we use these contours to generate
{$1\sigma$}-confidence allowed regions of {$N$(H)-$U$} for
components A1, B1, B2, B3, and C3. In Table~\ref{tab:rxjphyscond},
we report the optimal values for {$N$(H)} and {$U$} for components
B1, B2, and C2, and limiting constraints for components A1 and B3.
For component B2, {\SiIII} and {\SiIV} are overproduced by
{$\sim0.2$~dex}. That is, the column density limits (from \S3.1)
are not satisfied. There is not enough information provided by the
data to examine the reason for this. Possible reasons include an
artifact of Gaussian line-broadening, abundance variations, or
partial coverage effects.

In the same table, we also report the predicted {\HI} and {\OVI}
column densities based on the best-fit physical conditions. Using
these predicted column densities, we constrain the {\HI} and
{\OVI} coverage fractions of the components using the profile
fitter, fixing the column densities and kinematics. These are also
listed in the table. No low ionization species (e.g., \CII, \SiII)
were detected in these spectra. Such detections would provide
constraints on the density (in conjunction with the corresponding
strong high excitation lines), which would further constrain the
relation. More importantly, it would be prudent to obtain column
densities of higher ionization species to constrain the ionization
parameter. Using only {\NV} and {\CIV} column densities is
problematic when the ionization parameter is {$U\gtrsim0.2$},
since small uncertainties in the column densities propagate to
large uncertainties in {$N(\mathrm{H})$} and {$U$}.

\section{Summary of Results}
\label{sec:rxjresults}

The ultraviolet spectrum of {RX~J$1230.8+0115$} hosts a complex of
associated narrow absorption line systems detected in {\HI},
{\CIV}, {\NV}, and {\OVI}. Before detailed analysis, there are a
few striking properties of this complex to note. First, the four
systems reside in a {$5000$~\kms} velocity range that covers the
QSO emission redshift. System D lies redward of the broad emission
line peak, and only {$\sim50$~\kms} redward of the systemic
redshift. System B has very smooth trough indicative of a
wind-like outflow, a mini-BAL. The most striking thing to note is
that Systems A and C appear to be line-locked with the mini-BAL on
{\it opposite} sides in velocity space. System A appears to be
line-locked in the {\OVI} doublet at a larger ejection velocity,
while system C appears to be locked in the {\NV} doublet at a
smaller velocity.

Profile fits to the {\HI\,\Lya}, {\CIVdblt}, and {\NVdblt}
transitions available in the STIS-E140M spectrum of the four
systems, accounting for blends and coverage fractions, reveal that
Systems A, B, and C are likely to only partly occult the central
engine. They show coverage fractions less than unity. Column
densities derived from the same profile fits provide constraints
to the ionization conditions of the components. The data and
modelling procedure only allow the comparison of three components
in this sightline and, since most of the components have
ionization parameters {$U\gtrsim0.2$}, the {\NV} and {\CIV} column
densities do not provide stringent constraints. Although {\OVI}
information is available in the FUSE band, the quality of the
spectrum is not sufficient to disentangle the column densities and
coverage fractions of the components.

\section{Discussion}
\label{sec:rxjdiscuss}

We have provided strong evidence that the absorption complex
observed toward {RX~J$1230.8+0115$} is likely to arise from
intrinsic gas. Absorption with several components is fairly common
in lower-luminosity AGN (e.g., Seyferts) and broad absorption line
quasars show striking, continuous absorption over several thousand
{\kms} in velocity. However, narrow velocity-dispersion intrinsic
absorption in widely detached systems is uncommon. In the case of
this complex, an added curiosity comes from the coincidence
between absorber relative velocities and the separation of the
{\NVdblt} and {\OVIdblt} resonant doublets. We explore possible
origins of these absorbers assuming that coincident velocities are
due to physical line-locking.

Since the velocity separations between systems A, B, and C
coincide with the doublet spacing of {\NV} and {\OVI} doublet, we
focus on locking of absorption lines. There are two prevalent
dynamical approaches toward explaining absorption-absorption
line-locking: steady state in which the acceleration of absorbers
falls to zero and velocities remain constant
{\citep[e.g.,][]{milne26,scargle73}}; and time-variable in which
the {\it relative} acceleration between absorbers is zero leading
to a constant relative velocity {\citep{bm89}}. In both cases,
radiation pressure from the quasar central engine drives the
dynamics of the absorbers. We note here that if the absorbers are
truly line-locked, then the fact that we {\it observe}
line-locking implies we must be looking ``down the wind'' to avoid
projection effects. (Other orientations would result in observed
velocity differences smaller than the doublet spacing.)

Can the absorbers arise in line-locked steady state flow? If the
absorbers result from the instabilities described by Milne and
Scargle, then they must arise within the escape radius of the
black hole for gravity to provide a sufficient counter-force. The
escape radius is given by {$r = 0.03M_8v_{\mathrm{5k}}^{-2}$\,pc},
where {$M_8$} is the black hole mass in units of
{$10^8$\,M$_{\odot}$} and {$v_{\mathrm{5k}}$} is the absorber
velocity in units of {5000\,\kms}. This places very serious
constraints on the absorber densities, since we know their
ionization parameters. From the scaling laws derived from
reverberation mapping {\citep{kaspi00}}, the size of the BLR for
this quasar is {$\approx0.2$\,pc} and the black hole mass is {$6.2
\times 10^7~\mathrm{M}_{\odot}$}. These absorbers must then lie
within {$0.02$\,pc} of the continuum region, well within the BLR.
The implied densities are in excess of
{$10^{11}~\mathrm{cm}^{-3}$} with recombination timescales on the
order of seconds and absorber thicknesses less than
{$0.2~R_{\sun}$}. Since the absorbers suppress both continuum and
emission line photons, they must arise, at the very minimum,
co-spatially with the BLR. If the BLR extends down to
{$\sim0.01$\,pc} {\citep[see, for example, the simluation from
][]{mur95}}, then this scenario may be plausible, but requires
finely-tuned parameters.

Do the absorbers fit into the Braun \& Milgrom line-locking
(hereafter, BMLL) prescription? In this prescription, the absolute
velocity (with respect to the source of radiation) is not
important; only the velocity differences between parcels of
accelerated gas matter. BMLL takes advantage of the idea that
line-locking fundamentally occurs when two parcels of gas
experience the same acceleration, resulting in a constant velocity
{\it difference}. The two attractive features of BMLL are: (1)
there is no need for a counter force (e.g., gravity, or drag); and
(2) gas elements at several different velocities can be locked at
constant velocity differences. Braun \& Milgrom propose that a
time-variable (i.e., non-steady state) radiatively driven wind can
accomplish this. In the remainder of this discussion, we attempt
to incorporate BMLL into the accretion-disk/wind scenario
{\citep*[e.g., ][]{mur95,psk00}}.

Although, {\citet{bm89}} do not cite a cause for a time-variable
wind, we propose that such variability is to be expected given the
observed variability of quasar light curves {\citep[e.g.,
][]{kaspi00}}. By simple continuity arguments, the mass outflow
rate is regulated by the difference between the mass accretion
rate and the mass fuelling rate: $\dot{M}_{\mathrm{out}} =
\dot{M}_{\mathrm{fuel}} - \dot{M}_{\mathrm{acc}}$. The mass
accretion rate in turn is capped by the Eddington rate. For this
quasar, \citet{rmh98} report
{FWHM$_{\mathrm{\Hb}}\approx1500$\,\kms}, implying that the
Eddington rate is small ($\sim0.1~\mathrm{M_{\odot}~yr^{-1}}$).
The bolometric luminosity of the quasar is
{$\sim10^{45}~\lumunit$}, implying an accretion rate of
{$\sim0.15/\eta~\mathrm{M_{\odot}~yr^{-1}}$}, where {$\eta$} is
the mass accretion efficiency. Thus, the quasar is accreting near
its Eddington rate.

In Fig.~\ref{fig:rxjdiskwind}, we draw a possible cartoon of the
intrinsic absorbers within the accretion-disk/wind scenario. We
schematically label possible locations for systems A, B, and C. We
propose that a stochastic fluctuation in the mass outflow rate
(manifested either by a decrease in the mass accretion rate, or an
increase in the mass fuelling rate) results in an enhancement of
the wind density which propagates down the wind. We associate this
enhancement with system C. Due to radiative transfer effects
through system C, a region ``down-wind'' will see a decrease in
line pressure and become locked at the same acceleration as system
C. Gas in between these regions will continue to accelerate at a
rate greater than the locked region, resulting in an evacuation of
the `in-between'' region and a pile-up in the locked region. We
associate gas in this locked region with system B. By the same
arguments, system A is the result of gas that piles up in the
region down wind locked by system B. Given the clumpiness of the
system D's kinematics and lack of partial coverage, it is likely
that it arises from host galaxy (or nearby intervening) gas.

Several questions remain to be answered regarding this scenario.
\begin{itemize}
\item[1.] Is it feasible? Numerical simulations involving changes
in the mass outflow rate are needed to address this question. Even
so, current hydrodynamic codes invoking the Sobolev treatment may
not be sufficiently equipped to form this type of ``instability''
\citep[e.g., ][]{ro02}. Proper simulations will also show over
what timescales such instabilities form and how long they can
persist. In the case of {RX~J$1230.8+0115$}, these features were
observed in GHRS-G140L spectra {\citep{imp99}}, a persistence time
on the scale of at least 10 months in the QSO rest-frame.
\item[2.] How frequent is it? While it is clear that the majority
of absorption systems arise from intervening gas, as many as
{30\%} of {\CIV}--selected systems may be intrinsic to the QSO
{\citep{rich99}}. We note that it is possible that multiple
line-locked systems may not be observed to be line-locked given
velocity projection effects.
\item[3.] Does the scenario properly account for partial coverage
of the continuum and broad emission line regions? In the simple
two dimensional cut shown in the cartoon, it appears that half of
each of these regions is occulted, whereas the coverage fractions
from Table~\ref{tab:rxjintrin} imply otherwise. It is certain that
most of the continuum source is occulted. Thus it is likely that
the opening angle of the wind is very shallow and runs mostly
perpendicular to the disk axis (more than drawn). This has the
effect of allowing the absorbers to also occult the far side of
the disk (since the orientation is still one looking down the
wind).
\end{itemize}

Independent of the plausibility of the line-locking phenomenon,
the intrinsic absorption complex toward {RX~J$1230.8+0115$} merits
further study. A pointed observation with Chandra or XMM-Newton to
look for the presence and variability of ``warm'' absorption would
severely constrain the gas densities and photoionization
parameters. Similarly, follow-up observations in the ultraviolet
(with HST/STIS or HST/COS) to look for variability would constrain
the densities of the absorbers and the source-absorber distances.
Higher quality ultraviolet spectra, especially those covering
higher ionization lines, would allow better constraints on partial
coverage fractions and column densities. Observations to
separately constrain the ionization conditions and the densities
of the components yield direct constraints on the source-absorber
distances, and the absorber thickness {\citep[e.g., ][]{ham97b}}.
A further test of the scenario would come from a deep radio
observation to constrain the inclination of the disk, although it
may be possible to do so using broad emission lines
{\citep{rich02b}}.

\acknowledgements

We thank Michael Rauch for obtaining the STIS-E140M spectrum and
encouraging our work on the intrinsic absorption complex. The
NASA/ESA {\it Hubble Space Telescope} observations were carried
out at The Space Telescope Science Institute, which is operated by
the Association of Universities for Research in Astronomy, Inc.
under NASA contract NAS 5-26555. The FUSE spectra were obtained
for the FUSE Science Team by the NASA-CNES-CSA FUSE mission
operated by the Johns Hopkins University. We thank the members of
the Hubble Space Telescope Quasar Absorption Line Key Project for
providing a completely reduced, calibrated, and continuum-fit
dataset. We also thank W.~N. Brandt for assistance in assessing
and analyzing the X-ray observations of {RX~J$1230.8+0115$}. In
addition, we thank the anonymous referee for several incisive
comments leading to a much improved paper. This work was funded
through NASA grant NAG5-10817 at Penn State, and GO-08097.02A at
The Space Telescope Science Institute.

\appendix

\section{Accuracy of the Modelling Prescription}

In this appendix, we address the validity of applying
eq.~\ref{eq:cloudattenuation} in constraining the coverage
fractions and optical depths of absorption components in the limit
where such components can be treated as Gaussian-broadened. In the
text, we have already shown that this expression reduces to the
correct formulation in the following two limits: (1) the coverage
fractions are unity; and (2) component optical depth are large.

The simplest way to show that the prescription yields accurate
results in all other cases is to apply it to real profiles and
compare the resulting coverage fractions and optical depths to
those derived using the standard direct inversion technique
(eqs.~\ref{eq:atten} and \ref{eq:parcov}). However, in the absence
of a large data set of high spectral-resolution profiles of
intrinsic absorbers, we choose the following Monte Carlo approach:
(1) synthesize profiles using eq~\ref{eq:cloudattenuation}; (2)
apply the direct inversion technique to those profiles; and (3) at
the central wavelength of each component in the profile, record
the optical depth and coverage fraction derived from both methods.
In this way, we can control the range of optical depths and
coverage fractions via the input distributions.

In our simulations, we synthesized profiles using the {\CIVdblt}
doublet. The number of components in each profile was chosen
randomly on the interval {[1,5]}, with the velocity of each
component chosen on the interval {[-200,200]~\kms}. The coverage
fraction of each component was drawn uniformly on the range
{[0,1]}. The column density of each component was drawn from a
power law distribution on the interval
{[$10^{12.5}$,$10^{15.5}$]~$\mathrm{cm}^{-2}$}, with the
probability of drawing a column density between {$N$} and {$N+dN$}
being {$P(N)dN=CN^{-1}dN$}, where the normalization constant
{$C=1/\ln(10^{15.5}/10^{12.5})=0.1448$}. We chose a velocity width
distribution based on the {\CIV} ionization fraction distribution
expected if the gas were in collisional ionization equilibrium.
Using the {\CIV} ionization fraction versus temperature tabulated
by {\citet{sd93}}, we transformed the temperature to doppler width
($b^2=2kT/m$) and treated the (appropriately normalized)
ionization fraction as the probability {$P(b) db$} of drawing a
doppler width between {$b$} and {$b+db$}.

At the velocity {$v_j$} of each component, we compute the
``model'' optical depth {$\tau_j$} from the column densities
{$N_i$}, velocities {$v_i$}, and velocity widths {$b_i$} via:
\begin{equation}
\tau_j = {{\sqrt{\pi} e^2} \over {m_e c}} \sum_{i} N_i \exp \left
[ - \left ( {{v_j - v_i} \over {b}} \right )^2 \right ],
\end{equation}
where we perform a sum over all components to ensure a proper
comparison to the direct inversion optical depth computed with
eqs.~\ref{eq:atten} and {\ref{eq:parcov}}.

We synthesized {$10^6$} profiles resulting in {$3\times10^6$}
components to use in the comparison of our method with direct
inversion. In Fig.~\ref{fig:model}, we plot the distribution of
differences between the model and direct inversion values. From
the plots, it is clear that there is no systematic offset between
the model parameters and those derived through direct inversion
although there is a small tail toward
{$\tau_{\mathrm{model}}>\tau_{\mathrm{direct}}$} and
{$C_{\mathrm{model}}<C_{\mathrm{direct}}$}. Moreover, the
{$1\sigma$} systematic uncertainties in using our modelling
prescription are as follows: $\Delta C=0.00_{-0.01}^{+0.00}$,
$\Delta\tau=0.00_{-0.00}^{+0.06}$. We therefore conclude that, in
the limit where it is appropriate to model absorption profiles as
arising from discrete, Gaussian-broadened components that
partially occult the background source, our prescription
accurately reproduces both the coverage fraction and optical depth
(hence, the column density) of the components independent of the
geometric interpretation of the coverage fraction.

\clearpage

\begin{deluxetable}{lcrrrrrr}
%\tablenum{1}
%\tabletypesize{\small}
\tablewidth{0pc}
\tablecaption{Intrinsic Absorption}
\tablehead {
& & & \colhead{\HI} & \multicolumn{2}{c}{\CIV} &
\multicolumn{2}{c}{\NV} \\
%\cline{4-4}\cline{5-6}\cline{7-8}
%
& & & \colhead{\hrulefill} & \multicolumn{2}{c}{\hrulefill} &
\multicolumn{2}{c}{\hrulefill} \\
& \colhead{$\zabs$\tablenotemark{a}} & \colhead{$b$} &
\colhead{$\log~N$\tablenotemark{b}} &
\colhead{$\log~N$\tablenotemark{c}} & \colhead{$C$} &
\colhead{$\log~N$\tablenotemark{c}} &
\colhead{$C$} \\
\colhead{Comp.} & & \colhead{$\sigma_{b}$} &
\colhead{$\sigma_{\log~N}$} & \colhead{$\sigma_{\log~N}$} &
\colhead{$\sigma_{C}$} &
\colhead{$\sigma_{\log~N}$} & \colhead{$\sigma_{C}$} \\
 & & \colhead{(\kms)} & \colhead{(cm$^{-2}$)} & \colhead{(cm$^{-2}$)} &
& \colhead{(cm$^{-2}$)} &}
\startdata
A1 & 0.10012 &  55.1 & 12.91 & $>$14.4 & 0.14 & $>$14.6 & 0.24 \\
   &         &   5.9 &  0.10 & \nodata  & 0.02 & \nodata & 0.01 \\
\hline
B1 & 0.10529 &  78.9 & 14.09 & 14.71 & 0.58 & 14.89 & 0.78 \\
   &         &   3.3 &  0.01 &  0.14 & 0.03 &  0.04 & 0.01 \\
\cline{2-8}
B2 & 0.10603 & 164.9 & 13.97 & 14.97 & 0.49 & 14.66 & 0.95 \\
   &         &  13.9 &  0.02 &  0.12 & 0.03 &  0.23 & 0.15 \\
\cline{2-8}
B3 & 0.10639 &  38.9 & 13.67 & 14.23 & 0.34 & $>$14.5 & 0.59 \\
   &         &   3.1 &  0.00 &  0.50 & 0.09 & \nodata & 0.02 \\
\hline
C1 & 0.10929 &  28.3 & 13.40 &\nodata  &\nodata  &\nodata  & \nodata \\
   &         &   4.3 &  0.05 &\nodata  &\nodata  &\nodata  & \nodata \\
\cline{2-8}
C2 & 0.10937 & 101.1 & 13.69 & 14.24 & 1.00 & 14.90 & 0.89 \\
   &         &   4.9 &  0.03 &  0.19 & 0.14 &  0.04 & 0.02 \\
\hline
D1 & 0.11697 &  48.9 & 13.30 &\nodata  &\nodata  & 13.92 & \nodata \\
   &         &   0.9 &  0.03 &\nodata  &\nodata  &  0.03 & \nodata \\
\cline{2-8}
D2 & 0.11708 &  13.7 & 12.89 &\nodata  &\nodata  & 13.94 & \nodata \\
   &         &   0.6 &  0.06 &\nodata  &\nodata  &  0.03 & \nodata \\
\cline{2-8}
D3 & 0.11723 &  18.1 & 11.94 &\nodata  &\nodata  & 13.19 & \nodata \\
   &         &   3.2 &  0.45 &\nodata  &\nodata  &  0.15 & \nodata \\
\cline{2-8}
D4 & 0.11739 &  18.3 & 12.52 &\nodata  &\nodata  & 13.89 & \nodata \\
   &         &   0.6 &  0.11 &\nodata  &\nodata  &  0.03 & \nodata \\
\cline{2-8}
D5 & 0.11768 &  29.0 &\nodata  &\nodata  &\nodata  & 13.24 & \nodata \\
   &         &  12.1 &\nodata  &\nodata  &\nodata  &  0.11 & \nodata \\
\enddata
\tablenotetext{a}{The error in centroiding the redshift is
estimated at {$\sigma_{\zabs}\approx5$~\kms}.}
\tablenotetext{b}{The {\HI} column densities for systems A, B, and
C should be regarded as lower limits when the corresponding {\CIV}
and {\NV} components require partial coverage.}
\tablenotetext{c}{{$3\sigma$} column density limits are reported
when the optical depth (in the strongest available transition)
exceeds three.}
\label{tab:rxjintrin}
\end{deluxetable}

\begin{deluxetable}{crccccc}
%\tablenum{2}
\tablewidth{0pc}
\tablecaption{Physical Conditions of Components}
\tablehead {
\colhead{Name} & \colhead{$\log U$} & \colhead{$\log N(\mathrm{H})$} & \colhead{$\log N$(\HI)\tablenotemark{a}} & \colhead{$\log N$(\OVI)\tablenotemark{a}} & \colhead{$C$(\HI)\tablenotemark{b}} & \colhead{$C$(\OVI)\tablenotemark{b}} \\
               &                    & \colhead{$(\mathrm{cm}^{-2})$} & \colhead{$(\mathrm{cm}^{-2})$}           &
\colhead{$(\mathrm{cm}^{-2})$}         &                    & }
\startdata
A1 & \nodata  & $>19$ & \nodata & \nodata & \nodata   & \nodata \\
B1 & -0.45    & 19.8  & 14.8 & 16.2 & {$0.53\pm0.01$} & {$0.74\pm0.01$} \\
B2 & -1.28    & 19.1  & 15.0 & 15.3 & {$0.31\pm0.01$} & {$0.99\pm0.01$} \\
B3 & $>-1.15$ & $>19$ & \nodata & \nodata & \nodata   & \nodata \\
C2 &  0.33    & 21.0  & 15.0 & 16.5 & {$0.22\pm0.01$} & {$0.73\pm0.01$} \\
\enddata
\tablenotetext{a}{The {\HI} and {\OVI} were inferred from the
total column density and ionization paramter.}
\tablenotetext{b}{The coverage fractions for {\HI} were inferred
from the STIS data by fixing {\HI} column density at the Cloudy
prediction and tying the kinematics to the {\NV} and {\CIV}.
Similarly, the {\OVI} coverage fraction were inferred from the
FUSE data holding the column densities fixed at the inferred
values and tying the kinematics to the {\NV} and {\CIV} profiles
from the STIS data. Reported uncertainies are {$1\sigma$} (68\%)
confidence.}
\label{tab:rxjphyscond}
\end{deluxetable}

\clearpage

\begin{figure*}
%\vglue -0.7in
\figurenum{1}
\begin{center}
\epsscale{0.75}
%\rotatebox{270}{\plotone{f1.eps}}
\rotatebox{270}{\plotone{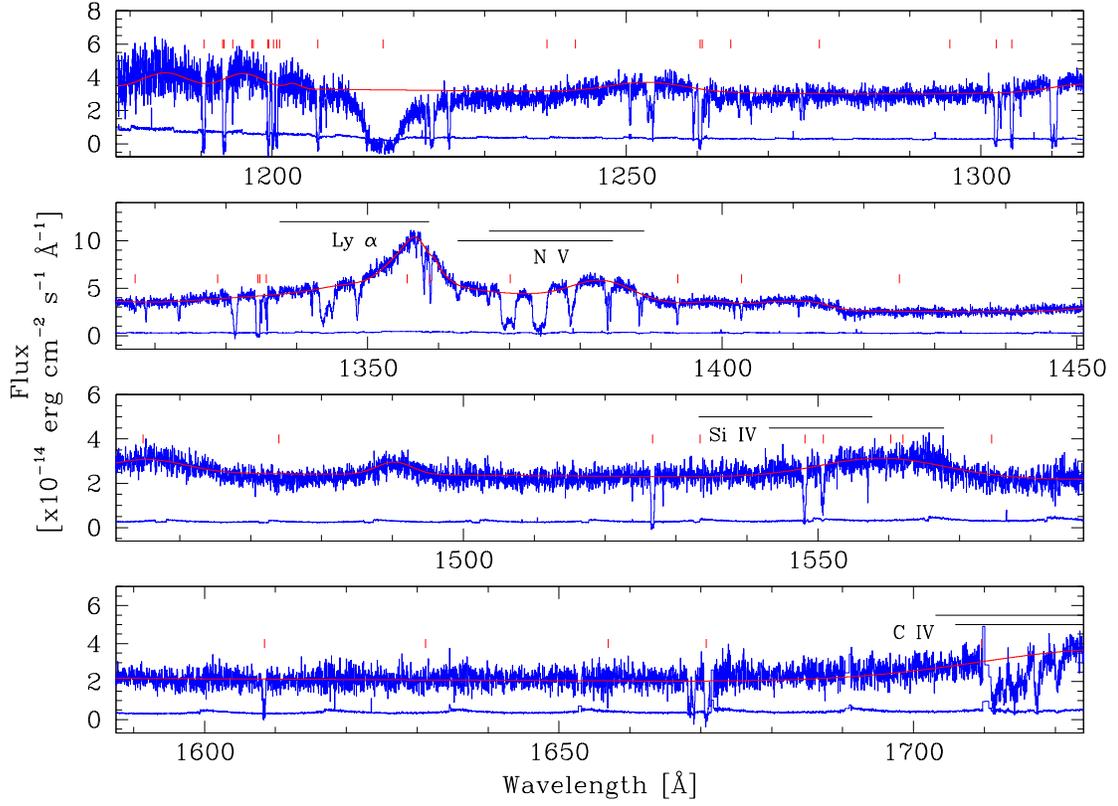}}
\end{center}
\caption[Fully calibrated HST/STIS-E140M spectrum of
RX~J$1230.8+0115$]{{\bf Fully calibrated HST/STIS-E140M echelle
spectrum of RX~J$\mathbf{1230.8+0115}$:} The spectrum covers the
wavelength range 1178.2--1723.8~\AA, with continuous coverage in
the region 1178.2--1634~\AA. The spectrum has been resampled and
rebinned to resolution element samples to reduce noise. The lower
trace in each panel is the {$1\sigma$} error spectrum. The
wavelength ranges covered by associated absorption are shown for
{\Lya}, {\NVdblt}, {\SiIVdblt}, and {\CIVdblt}. Vertical ticks
mark the expected locations of Galactic absorption from {\HI},
{\CI}, {\CII}, {\CIV}, {\NI}, {\NV}, {\OI}, {\MgI}, {\AlII},
{\SiII}, {\SiIII}, {\SiIV}, {\PII}, {\PIII}, {\SI}, {\SII},
{\SIII}, {\MnII}, {\FeII}, {\NiII}, and {\CuII}.}
\label{fig:rxjstisplot}
\end{figure*}

%\clearpage

\begin{figure*}
\figurenum{2}
%\vglue -0.7in
\begin{center}
\epsscale{0.75}
%\rotatebox{270}{\plotone{f2.eps}}
\rotatebox{270}{\plotone{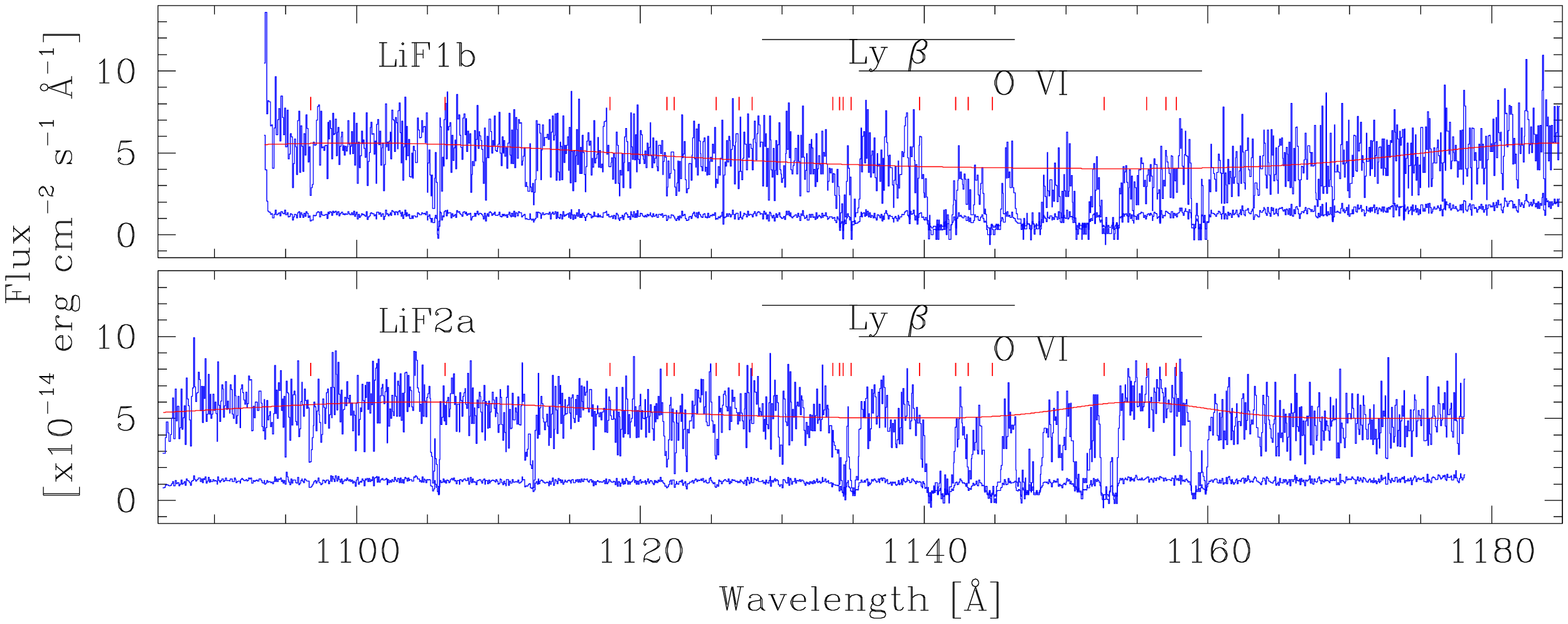}}
\end{center}
\caption[Fully calibrated FUSE-LiF spectrum of
RX~J$1230.8+0115$]{{\bf Fully calibrated FUSE-LiF spectrum of
RX~J$\mathbf{1230.8+0115}$:} In the above plot, we show spectra
from the two FUSE channels (LiF1b and LiF2a) which cover the
{\OVIdblt} and {\Lyb} transitions for the associated absorbers.
The spectra cover the wavelength range 1086.4--1184.8~\AA. The
spectrum has been resampled and rebinned to resolution element
samples to reduce noise. The lower trace in each panel is the
{$1\sigma$} error spectrum. The wavelength ranges covered by
associated absorption are shown for {\Lyb}, and {\OVIdblt}.
Vertical ticks mark the expected locations of Galactic absorption
from {\CI}, {\NI}, {\PII}, and {\FeII}.} \label{fig:rxjfuseplot}
\end{figure*}

%\clearpage

\begin{figure*}
\figurenum{3}
%\vglue -1.5in
\begin{center}
\epsscale{0.8}
%\plotone{f3.eps}
\plotone{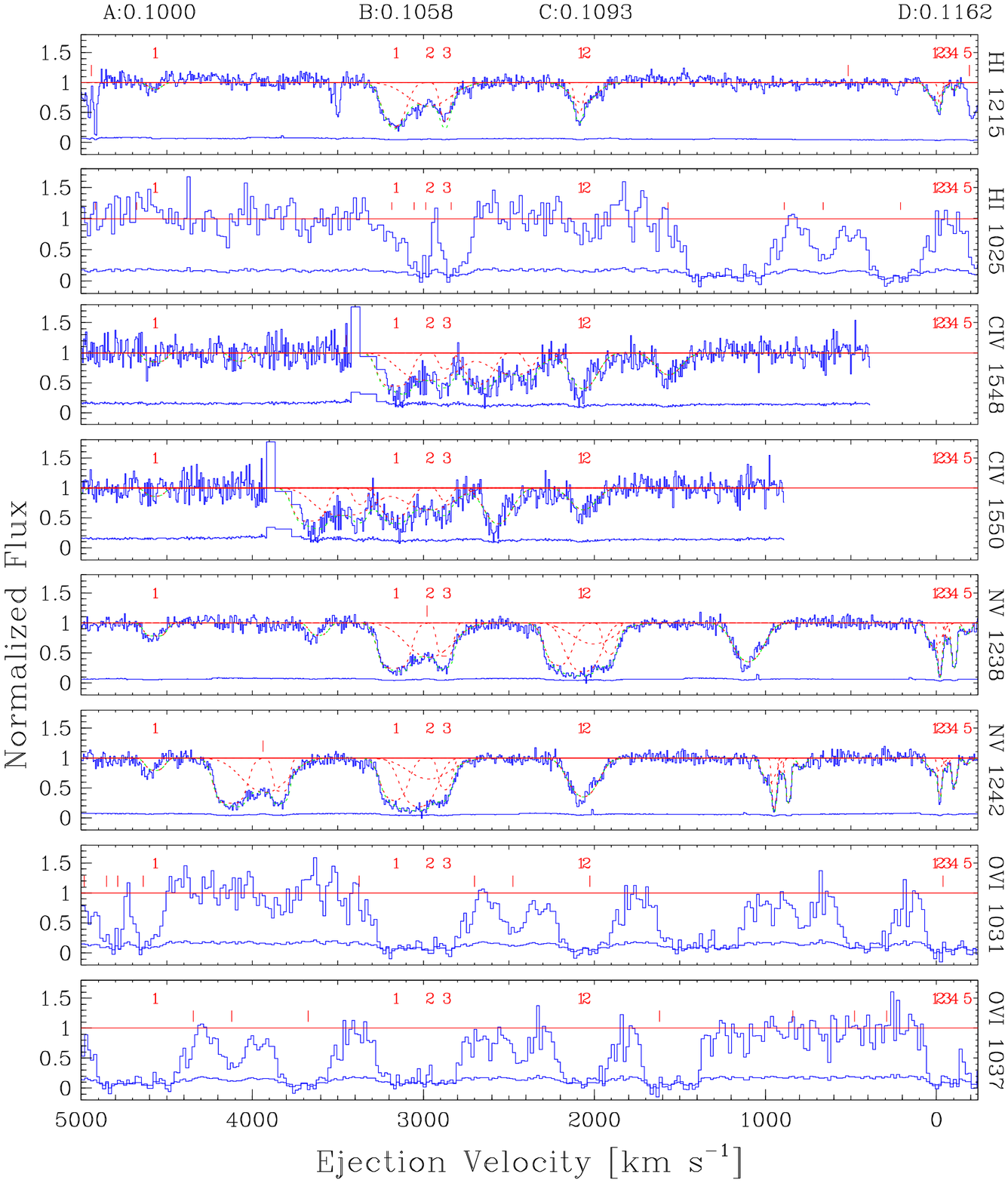}
\end{center}
\caption[Velocity Plot of {RX~J$1230.8+0115$} Intrinsic Absorption
Complex]{{\bf Velocity Plot of Absorption Complex:} The intrinsic
absorption complex is detected in {\HI~\Lya}, {\HI~\Lyb},
{\NVdblt}, {\CIVdblt}, and {\OVIdblt}. In this plot, we transform
the spectra to velocity space using {$z=0.117$} as the zero-point
for each of the five rest-wavelengths. The systems are labelled
A,B,C, and D as described in the text with their centroid
redshift. Within each system, we assign a component number. The
{\HI~\Lyb}, and {\OVI} absorption are data from the FUSE LiF2a
channel. The dotted curves in the STIS data ({\HI~\Lya},
{\NVdblt}, {\CIVdblt}) show the components used to fit the
absorption. For purposes of clarity, the Galactic {\NiII}
absorption has been removed from the {\NV} absorption in system
B.} \label{fig:rxjvelocity}
\end{figure*}

%\clearpage

\begin{figure*}
\figurenum{4}
\begin{center}
\epsscale{0.75}
%\rotatebox{270}{\plotone{f4.eps}}
\rotatebox{270}{\plotone{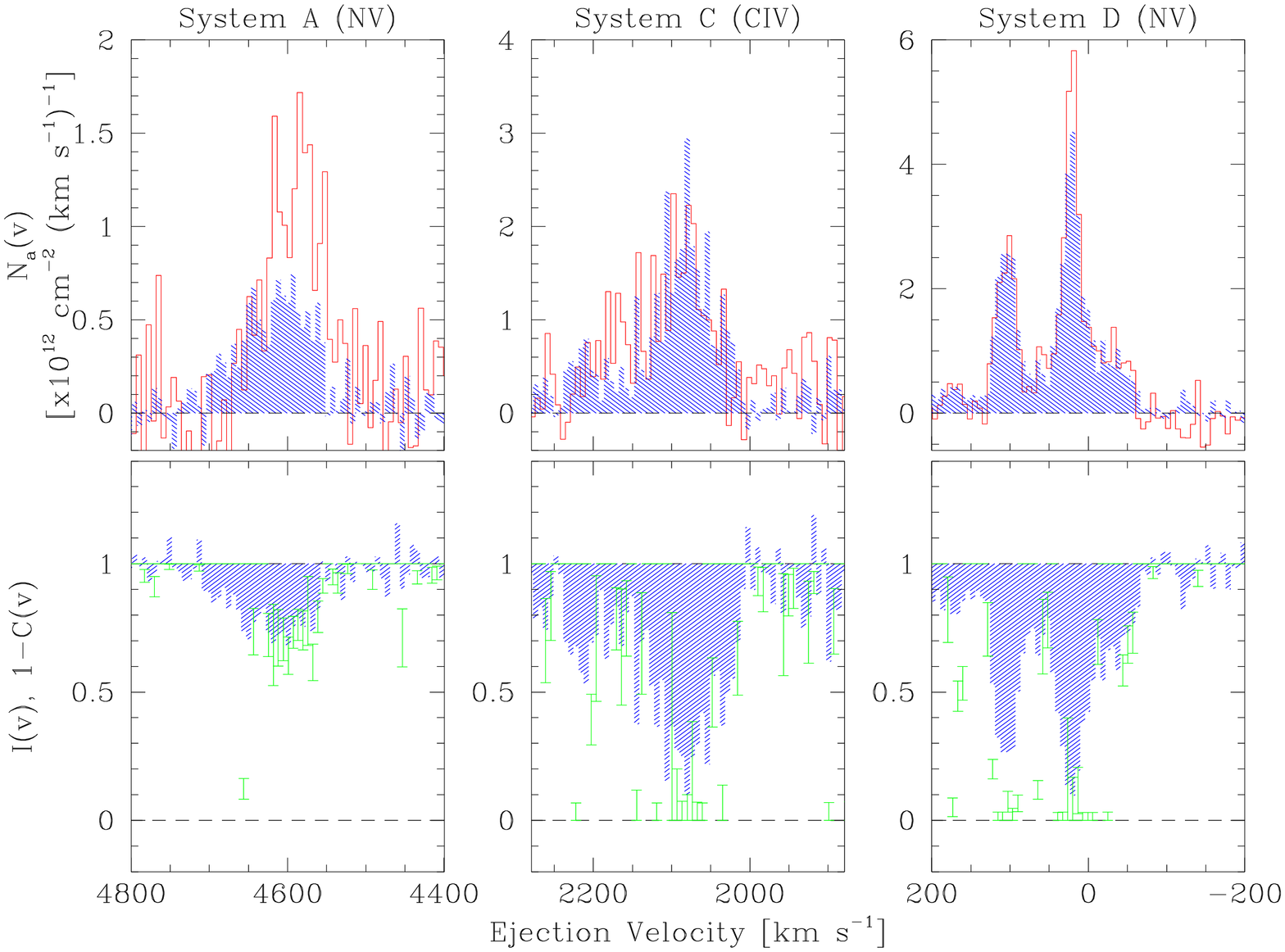}}
\end{center}
\caption[Partial Coverage Plots]{{\bf Partial Coverage Plots:} In
the above panels, we directly invert the unblended absorption
systems to test for the presence of diluted profiles. This is done
for doublet from three systems: {\NVdblt} for systems A (left) and
D (right), and {\CIVdblt} for system C (middle). The top panels
show the velocity-aligned apparent column density profiles from
the two transitions (stronger as shaded histogram, weaker as
unshaded histogram). In the limit of fully resolved profiles, the
signature of dilution is higher predicted column densities from
the weaker transition. In the bottom panels we show the normalized
flux from the stronger transition (shaded histogram), and directly
compute the coverage fraction using eq.~\ref{eq:parcov} (shown
with error bars). Note that the coverage fractions are plotted on
an inverted scale to show the ``true'' limit of saturation.}
\label{fig:rxjparcov}
\end{figure*}

%\clearpage

\begin{figure*}
\figurenum{5}
\begin{center}
\epsscale{0.75}
%\rotatebox{270}{\plotone{f5.eps}}
\rotatebox{270}{\plotone{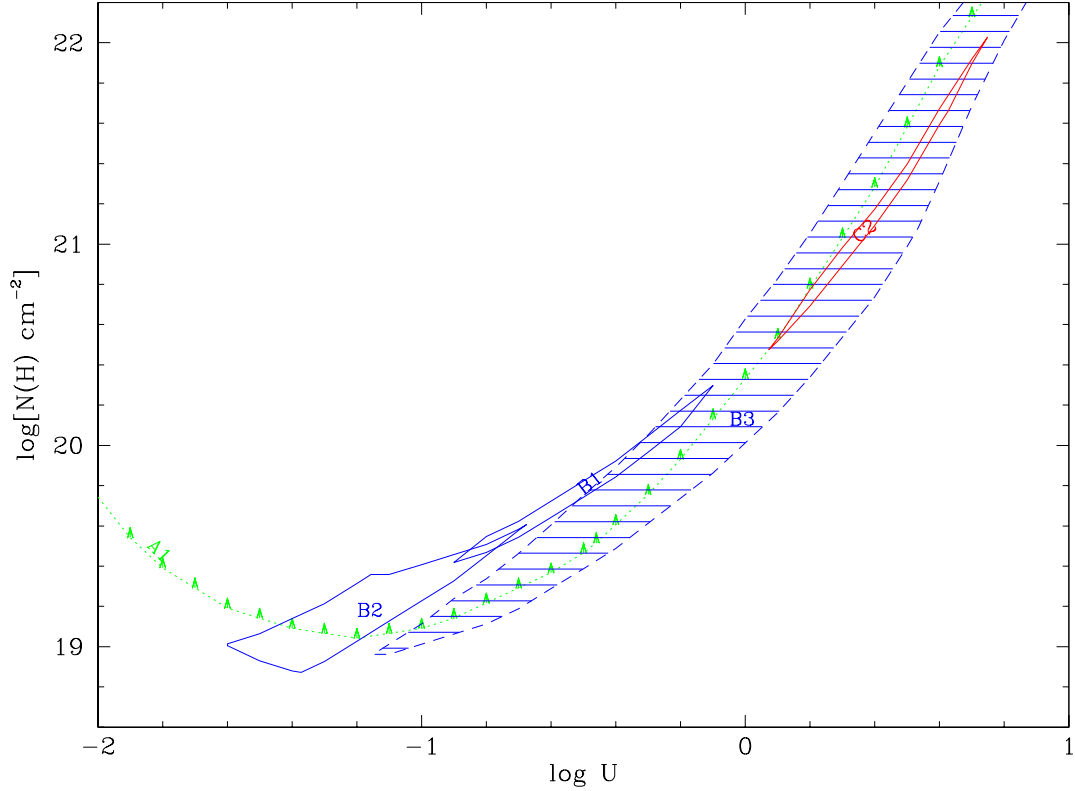}}
\end{center}
\caption[Ionization Conditions of Absorbers]{{\bf Ionization
Conditions of Absorbers:} We plot the allowed region ($1\sigma$
confidence) of total column density {$N(\mathrm{H})$} and  {$U$}
for each component with measured or constrained {\CIV} and {\NV}
column densities. For components B1, B2, and C2, enclosed regions
are shown as solid lines. For component B3, the allowed (hatched)
region continues off the figure since there is no upper constraint
on the {\NV} column density. Similarly, component A1 has only
lower limits for both {\NV} and {\CIV} column densities propagate
into a lower limit contour (dotted curve), with no upper bound.}
\label{fig:rxjionize}
\end{figure*}

%\clearpage

\begin{figure*}
\figurenum{6}
\begin{center}
\epsscale{1.0}
%\plotone{f6.eps}
\plotone{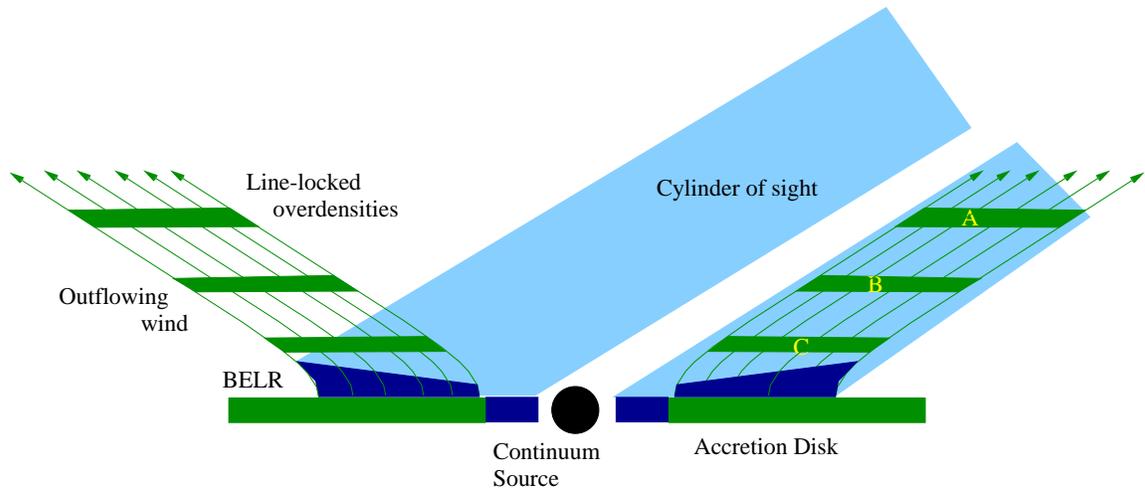}
\end{center}
\caption[Modified Accretion-Disk/Wind Cartoon]{{\bf Modified
Accretion-Disk/Wind Cartoon:} In this modification of the ``standard''
accretion-disk/wind model {\citep[e.g.,][]{mur95,psk00}}, we propose
that changes in the mass outflow rate can produce instabilities not
accounted for by current numerical simulations.  An instability at the
base of a line-driven wind may create line-locking instabilities at
the appropriate velocities. In the plot, we exaggerate the regions of
the wind where locked matter may pile-up. Dark (blue) regions indicate
the continuum and broad emission lines regions. All lines of sight
that reach the observer are shown in light shade (light blue). To
properly account for partial continuum source occultation, the wind
likely flows in a mostly radial direction.} \label{fig:rxjdiskwind}
\end{figure*}

\begin{figure*}
\figurenum{7}
\begin{center}
\epsscale{0.7}
%\rotatebox{-90}{\plotone{f7.eps}}
\rotatebox{-90}{\plotone{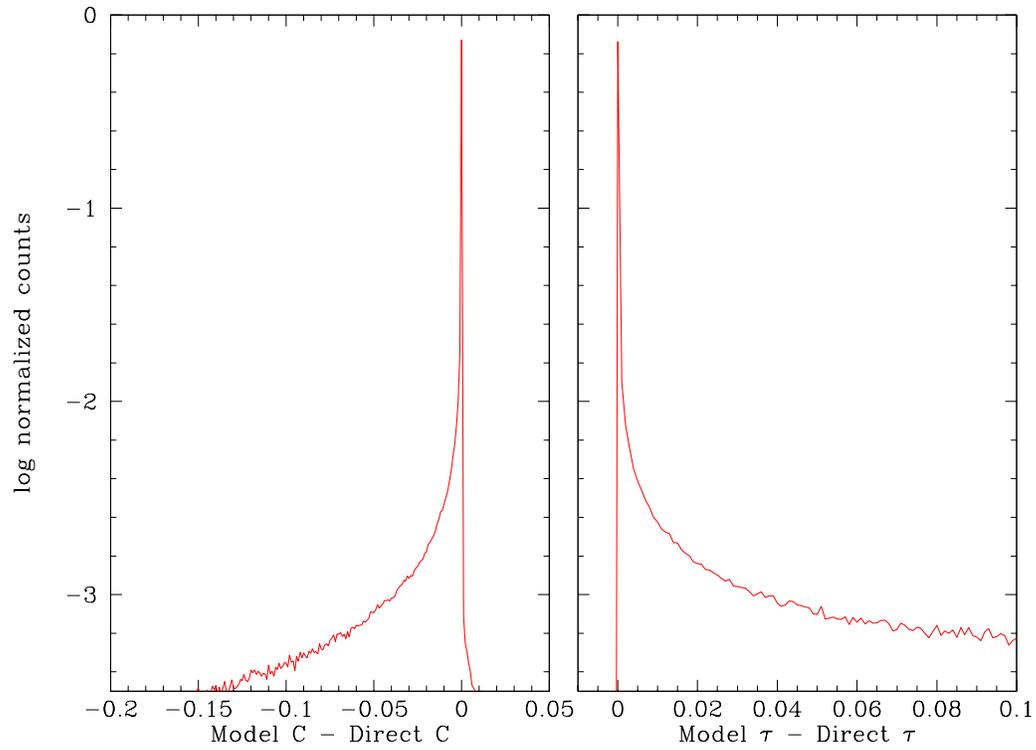}}
\end{center}
\caption[Accuracy of Model-Derived Coverage Fractions and Optical
Depths]{{\bf Accuracy of Model-Derived Coverage Fractions and
Optical Depths:} In the above panels, we show the histograms of
parameter differences (coverage fraction in the left panel;
optical depth in the right panel) between the model
(eq.~\ref{eq:cloudattenuation}) and the direct inversion
technique. The distributions peak at zero, indicating that there
is no systematic offset between the model and direct inversion.}
\label{fig:model}
\end{figure*}

\end{document}